# Characteristic Tip–Substrate Capacitance Studied Using Force Spectroscopy Method of Atomic Force Microscopy

*Reynier I. Revilla[§], Yan–Lian Yang[*], Chen Wang[*]*

Key Laboratory of Standardization and Measurement for Nanotechnology, the Chinese Academy of Sciences, National Center for Nanoscience and Technology, Beijing 100190, P. R. China

§ Present address: Center for Advanced Studies of Cuba, Havana 19370, Cuba

*Corresponding authors: Tel: 86-10-82545559, Fax: 86-10-62656765
Email: yangyl@nanoctr.cn (YLY); wangch@nanoctr.cn (CW)

**ABSTRACT:** The characteristic tip–substrate capacitance is crucial for understanding the localized electrical properties in atomic force microscopy (AFM). Since it is highly dependent on tip geometrical features, estimation of the tip–substrate characteristic capacitance for a given probe is very complex, involving empirical measurements and numerical simulations. In this paper we propose a facile approach to study the tip–substrate characteristic capacitance using AFM force spectroscopy technique. In this scheme, an analytical expression is considered to model the tip–sample interaction, in which the coefficients are directly dependent on the tip–substrate capacitance. This method avoids any complex simulation involving irregular shape of AFM tips. Additionally, it considerably reduces amount of experimental data needed for the calculation compared with other techniques. The work presented here also corroborates that for tip–sample separation lower than 200 nm, the parallel plate capacitor approximation is not very appropriate to describe the tip–substrate capacitive interaction, but an intermediate approximation between a sphere–plane and a cone–plane geometry seems to be more appropriate.

## I. INTRODUCTION

Electric force microscopy (EFM) and Kelvin probe force microscopy (KPFM) are widely used and very powerful techniques for characterizing electrical behaviors of nanostructures.[1] By using EFM and directly imaging the surface potential variations at nanoscale, valuable information have been obtained regarding the nature of charge transport,[2–8] conductivity,[9,10] and charge injection[8,11–13] in a great number of systems/materials. Nevertheless, in order to carry on quantitative measurements of dielectric properties such as charge density and dielectric permittivity, surface potential imaging alone is not enough. This is principally because the tip–surface electrostatic interaction includes contributions from several sources, for instance, tip–substrate capacitance, static charge at interfaces, boundaries and/or defects, remnant polarization, and so on. To facilitate quantification techniques, extensive efforts have been put forth to model tip–substrate capacitance for the interpretation of EFM images and measurements. This characteristic capacitance is highly dependent on geometrical features. Commercial EFM probes are of great complexity, *i.e.* an integrated structure including an approximately spherical tip apex, a pyramidal/conical tip body and a plate cantilever.[14] Several approaches have been used in the literature to model the scaling behavior of the tip–substrate characteristic capacitance. The parallel–plate capacitor geometry ($\partial^2 C/\partial z^2 \sim z^{-3}$) is one of the most extensively used due to its simplicity.[15,16] However, the power law of $z^{-3}$ could not describe most of the experiments results due to the discrepancy between the parallel–plate capacitor model and the practical tip–substrate geometry. In order to solve this problem, probe–plane characteristic capacitance ($\partial^2 C/\partial z^2 \sim z^{\beta}$) is proposed,[17,18] while a universal value ($\beta$) to describe the power law dependence is still under debate. For certain range of tip–sample distance, experimental results have shown that the tip–substrate capacitance lies intermediately between a sphere–plane ($\partial^2 C/\partial z^2 \sim z^{-2}$) and a cone–plane ($\partial^2 C/\partial z^2 \sim z^{-1}$) geometry ($\beta$ from –1.4 to –1.6),[18–20] which is in good agreement with an approximately conical tip with an spherical ending. It has been previously discussed that the apex and conical portions of an EFM tip are responsible for the greater part of the interactions when the tip is not far away from the sample surface, and the contribution from the cantilever plate becomes more appreciable at larger tip–sample distances.[21] Other researchers have also considered the total tip–sample capacitance as the sum of all the contributions from the different parts of the EFM probe.[20,22,23] Nevertheless, to estimate the characteristic capacitance for a given probe, complex procedures are usually required involving empirical measurements of capacitive force characteristics combined with numerical simulations.

Qi *et al.*[24] developed a method to determine the dependence of the tip–substrate capacitance on the tip–sample separation based on the effect of the sample surface bias. The characteristic capacitance could be determined by a single integration step, avoiding the introduction of additional linear terms as reported in previous studies.[22] This method is of great convenience when the experiments are performed on a conductive sample. If a dielectric film is deposited on the substrate, high values of the substrate bias could induce polarization within the film and therefore introduce an extra term in the interaction between the tip and the sample, which makes the interpretation of the data more complicated.

In this paper we present a simple method to study the scaling behavior of the characteristic capacitance between a conductive AFM tip and a substrate covered with a dielectric film. Force spectroscopy technique in combination with an analytical expression modeling the tip–substrate interaction forces is used to experimentally obtain a series of parameters directly dependent on the tip–substrate capacitance. In this way, one can further calculate the characteristic power–law of the system's capacitance with any commercially available AFM system without considering the tip geometry effect.

## II. EXPERIMENTAL SECTION

*AFM measurement*

Force spectroscopy measurements were conducted in ambient conditions with a commercial atomic force microscope (Dimension 3100, Nanoman II, Bruker). A rectangular conductive cantilever with a coating of a 3 nm Cr and 20 nm Pt–Ir film and a nominal spring constant of 3 N/m was used for the measurements and as the top electrode. The conductive substrate supporting the dielectric film was used as the bottom electrode.

In order to determine the average force curve for each external voltage applied to the tip, a series of 5 force curves were obtained on the same sample location (the substrate was always grounded). To avoid structural damage of the tip and the sample, the loading/unloading speed and the maximum cantilever deflection were kept relatively low, 0.25 μm/s and less than 40 nm respectively. The probe radius was measured before and after each test in order to ensure that the tip apex did not change due to plastic deformation. This control check was carried out using the NanoScope Analysis software and a standard Tip Check sample (Aurora NanoDevices Inc.), which are commonly used to obtain accurate reconstructions of the tip apex. Measurements were made at a relative humidity below 20%, in order to reduce the possible influence from the condensation of water on the sample surface.

*Sample preparation*

An 80 nm thick film of silicon dioxide was deposited on a silicon surface ($SiO_2$/Si(100)) by using physical vapor deposition (PVD). Si(100) was highly doped with phosphorus (2–4 Ω·cm) (from Silicon Quest Int'l) and used as the bottom electrode. PVD was carried out by E–beam deposition on a BOC AUTO 500 system, with a base vacuum of $1.6 \times 10^{-6}$ mbar, E–beam gun electric current of 7 mA and a deposit rate of 0.6 Å/s.

## III. RESULTS AND DISCUSSION

In previous works, we have proposed an analytical expression to model the total force acting on a biased EFM tip, $F_V(z)$, placed on top of a conductive substrate covered with a dielectric film.[25, 26]

$$F_V(z) = \frac{1}{2}\left[\frac{\partial C}{\partial z} + a(z)\chi_s C\right]V^2 + \frac{1}{2}a(z)\chi_s Q_{im}V + F_0(z) \tag{1}$$

Where $F_0(z)$ is the tip–sample interaction at zero voltage, which is principally associated with van der Waals interaction. $C$ is the effective tip–substrate capacitance and $z$ is the effective tip–substrate distance, thus $\partial C/\partial z$ is the first derivative of the tip–substrate capacitance with respect to the tip–substrate distance. $a(z)$ is a factor related only to the tip–sample distance and tip geometry, $\chi_s$ is the sample susceptibility and $Q_{im}$ the image charge on the conductive tip induced by the static charge in the sample below the tip apex.

According to the procedure described in our earlier works[25, 26] we can associate the coefficients of Eq. (1) to the functions $g(z)$ and $h(z)$ so that:

$$g(z) = \frac{\partial C}{\partial z} + a(z)\chi_s C \tag{2}$$

and

$$h(z) = a(z)\chi_s Q_{im} \tag{3}$$

These two functions can be directly obtained by measuring force–distance curves at 0 V and two other different voltages applied between the probe and the substrate.[25, 26]

From Eqs. (2) and (3), Eq. (4) can be easily derived:

$$\frac{\partial C}{\partial z} + \frac{1}{Q_{im}} h(z)C = g(z) \tag{4}$$

Eq. (4) is a first–order linear differential equation, which can be solved using a numerical integration method. In order to solve Eq. (4), we have considered the following initial condition, based on the fact that the tip–sample interaction force, and therefore functions $g(z)$ and $h(z)$, are approximately 0 for large tip–sample separation.

$$\frac{\partial C}{\partial z}\bigg|_{(z \gg 0)} = 0 \tag{5}$$

The parameter $Q_{im}$ in Eq. (4), which represents the image charge on the tip induced by the static charge on the sample surface below the tip apex, is generally unknown. Therefore, to resolve the differential equation represented below, we have assigned values to $Q_{im}$ ranging from 4 to approximately $10^3$ elemental charges (e⁻), and solve Eq. (4) for each one of those values given to $Q_{im}$. In other words, we have carried out a sensitivity analysis to determine how different values of the parameter $Q_{im}$ will impact the scaling behavior of the probe–sample characteristic capacitance.

Fig. 1(a) shows functions $g(z)$ and $h(z)$ obtained from force–distance curves measured on the SiO$_2$ film at 0, 3, and 6 V. These two functions are approximately 0 for relatively large tip–sample separation as mentioned above, which confirms the use of the boundary value represented in Eq. (5). Fig. 1(b) portrays the numerical solution of Eq.(4) for a selection of values assigned to the parameter $Q_{im}$ using the functions $g(z)$ and $h(z)$ portrayed in Figure 1(a).

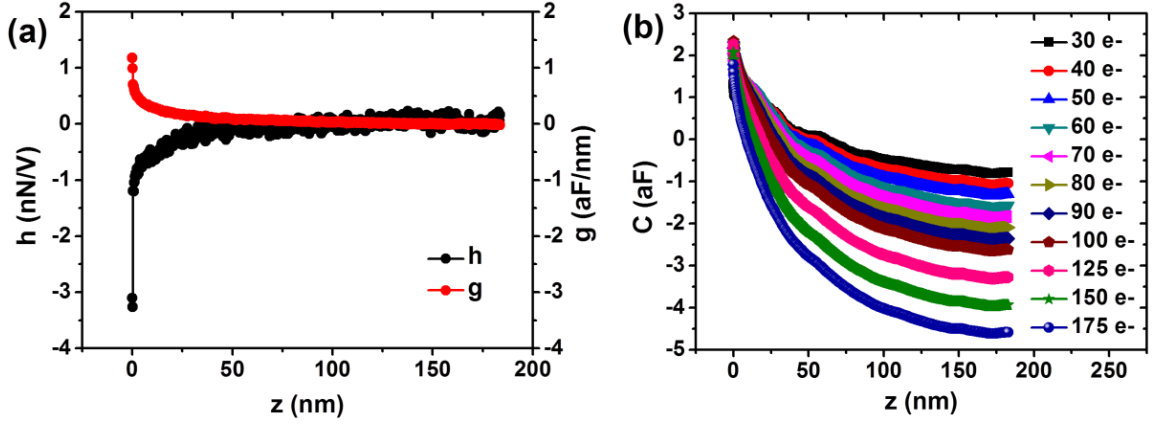

**FIG. 1.** (a) Value of $g(z)$ and $h(z)$ versus tip–sample distance, obtained from force–curves measured on a $SiO_2$ film at 0, 3, and 6 V applied between the probe and the substrate. (b) Numerical solution of Eq. (4) using functions $g(z)$ and $h(z)$ portrayed in (a), for a selection of values assigned to the variable $Q_{im}$.

In a previous work by Qi et al.[24] an analytical expression of the tip–substrate capacitance was proposed to study its scaling behavior:

$$C = A_1(z + z_{eff})^\alpha + A_2 \qquad (6)$$

Here $A_1$, $A_2$ and $z_{eff}$ are constants, where $z_{eff}$ is the parameter to correct the tip–sample distance $z$ to accommodate the effective charge position, and $\alpha$ is associated with the geometrical characteristics of the probe. It can be easily noted from Eq. (6) that the parameter $\beta$ mentioned above, which is associated with the power coefficient of the $z$–dependence of the second derivative of the capacitance, is related to parameter $\alpha$ through the relation: $\beta = \alpha - 2$. In order to study the scaling behavior of the tip–sample capacitance, the capacitance curves obtained from Eq. (4) were fitted using Eq. (6). Consequently, for every value of the parameter $Q_{im}$, one value of the power coefficient $\alpha$ can be obtained. Fig. 2 shows the values of $\alpha$ versus $Q_{im}$. The data fitting with Eq. (6) was conducted for values of $z$ ranging from 30 to 180 nm approximately. Values of tip–sample separation lower than 30 nm were ignored because when the probe is relatively close to the sample surface, this gets into an unstable regime in which the tip jumps into contact with the sample.[25] It can be clearly observed in Fig. 2 that coefficient $\alpha$ experiences almost no variation for $Q_{im}$ ranging from 10 to $10^3$ e⁻ approximately. Hence, $\alpha$ is relatively independent of the value of $Q_{im}$, which is in complete agreement with the fact that $\alpha$ is mainly related to the probe geometrical features. Values of $\alpha$ obtained during the fitting are principally in the range from 0.33 to 0.41. This range is represented by the shaded region in Fig. 2. Thus, the average value of $\alpha$ could be estimated as $\alpha = 0.37 \pm 0.04$, from which we can obtain a scaling behavior of the second derivative of the tip–substrate capacitance $\partial^2 C/\partial z^2 \sim z^{-1.63}$ ($\beta = -1.63$). This value lies intermediately between the sphere–plane and the cone–plane geometry. Consequently, this could indicate that the parallel plate capacitor approximation is not very appropriate to describe the tip–substrate capacitive interaction for tip–sample separation lower than 200 nm, which is the main range of operation of electric force microscopy. This also confirms that when the tip is not far away from the sample surface the apex and conical portions of the tip are responsible for most part of the interactions.

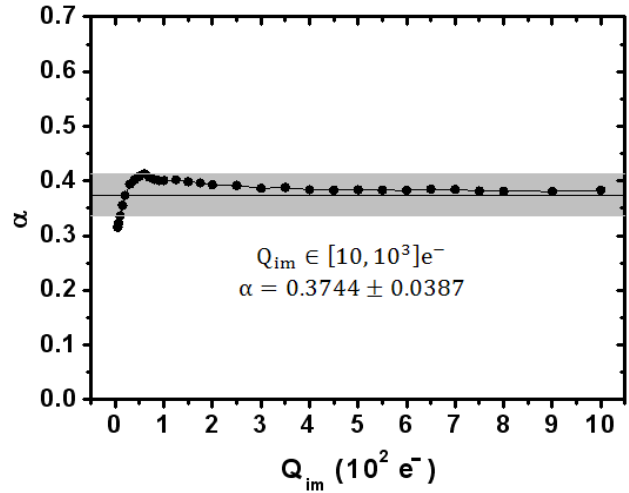

**FIG. 2.** Parameter $\alpha$ obtained from the fitting of the capacitance curves acquired from the numerical solution of Eq. (4) for different values of the parameter $Q_{im}$. The shaded region represents values of $\alpha$ from 0.33 to 0.41.

To further confirm our results, the same measurement was carried out using six different probes taken from two different batches with different cantilever' geometries (Batch #1: rectangular conductive cantilever with a coating of Pt–Ir/Cr, nominal radius 20 nm, nominal full tip cone angle 40°. Batch #2: V–shape conductive cantilever with a coating of Au/Cr, nominal radius 35 nm, nominal full tip cone angle 40°). For each probe the measurement was conducted twice. Fig. 3 portrays the values of the power coefficient obtained using the procedure explained above. It can be clearly observed that the values of $\alpha$ are within the range from 0.36 to 0.5, giving a scaling behavior of the second derivative of the characteristic capacitance from $z^{-1.5}$ to $z^{-1.64}$. This further confirms that for tip–sample distance ranging from 30 to 200 nm the main capacitive interaction comes from the tip apex and conical part. The cantilever contribution could almost be neglected in this case.

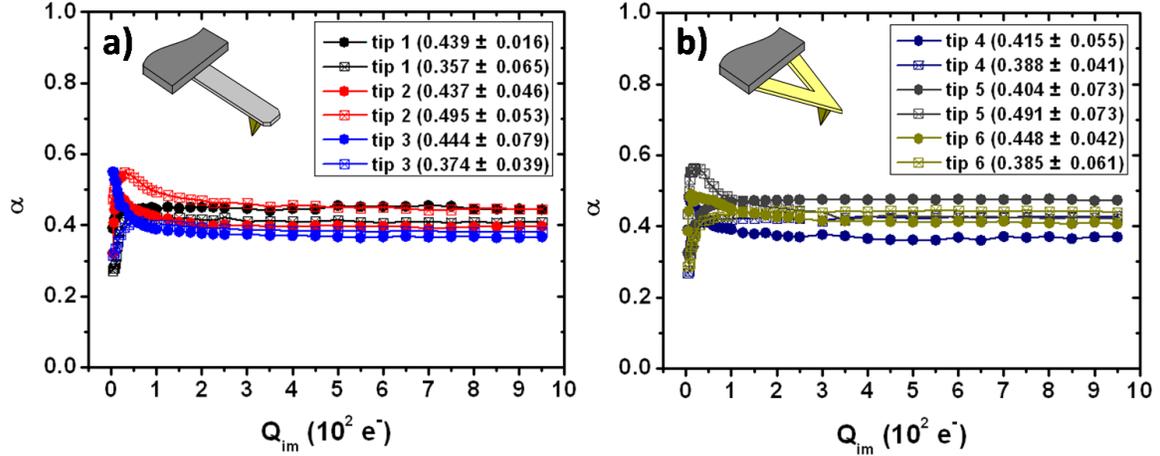

**FIG. 3.** Values of the coefficient $\alpha$ obtained using six different probes taken from two different batches (a) Batch #1: rectangular conductive cantilever with a coating of Pt–Ir/Cr, nominal radius 20 nm, nominal full tip cone angle 40º. (b) Batch #2: V–shape conductive cantilever with a coating of Au/Cr, nominal radius 35 nm, nominal full tip cone angle 40º.

The procedure described above estimates the scaling behavior of the absolute value of the probe–sample capacitance without requiring an integration constant as reported in previous works. This additional constant has usually been considered a limitation, requiring the evaluation of capacitance variations with respect to a given capacitance value instead of the absolute characteristic capacitance.[22] By knowing this scaling behavior a more realistic, yet simple, model to describe the electrostatic interaction between the probe and the sample can be used, avoiding the use/estimation of geometrical parameters of the structurally complex AFM probe.

Additionally, if a dielectric material is deposited between the biased probe and the substrate, an accurate analytical expression to model the tip–sample interaction force is extremely difficult to achieve. This is due to the introduction of an additional term in the interaction force associated with the induced polarization within the dielectric film. However, these systems (biased AFM probes – dielectric films) are widely used for electrical material characterization. In this regard the model proposed here represents an easy and convenient approach to obtain the feature of the probe–dielectric–substrate capacitive interaction for further quantification and study of other electrical properties.

## IV. CONCLUSIONS

We have proposed in this work a facile approach to study the scaling behavior of the tip substrate characteristic capacitance by using AFM force spectroscopy technique. This method avoids any complex simulation involving irregular shape of conductive AFM tips and considerably reduces the amount of experimental data needed for the calculation compared with other approaches. The work presented here also confirms that for tip–sample separation lower than 200 nm, the parallel plate capacitor approximation is not very appropriate to describe the tip–substrate capacitive interaction, but it seems more suitable a scaling behavior intermediate between the sphere–plane and the cone–plane geometry.

**ACKNOWLEDGMENT**

Financial support from National Key Basic Research Program of China (Grant Nos. 2013CB934200) and the National Natural Science Foundation of China (Grant Nos. 21273051) are gratefully acknowledged.

**REFERENCES**

[1] X. H. Qiu, G. C. Qi, Y. L. Yang and C. Wang, J. Solid State Chem. **181**, 1670 (2008).

[2] M. Zdrojeka, T. Melin, H. Diesinger, D. Stievenard, W. Gebicki and L. Adamowicz, J. App. Phys. **100**, 114326 (2006).

[3] M. Paillet, P. Poncharal and A. Zahab, Phys. Rev. Lett. **94**, 186801 (2005).

[4] C. A. Rezende, R. F. Gouveia, M. A. da Silva and F. Galembeck, J. Phys.: Condens. Matter **21**, 263002 (2009).

[5] T. A. de Lima–Burgo, C. Alves–Rezende, S. Bertazzo, A. Galembeck and F. Galembeck, J. Electrostat. **69**, 401 (2011).

[6] R. Wang, S. Wang, D. D. Zhang, Z. Li, Y. Fang and X. H. Qiu, ACS Nano **5**, 408 (2011).

[7] Z. Hu, M. D. Fischbein and M. Drndić, Nano Lett. **5**, 1463 (2005).

[8] C. Y. Ng, T. P. Chen, H. W. Lau, Y. Liu, M. S. Tse, O. K. Tan and V. S. W. Lim, Appl. Phys. Lett. **85**, 2941 (2004).

[9] D. Porath, A. Bezryadin, S. de Vries and C. Dekker, Nature **403**, 635 (2000).

[10] H. Cohen, T. Sapir, N. Borovok, T. Molotsky, R. D. Felice, A. B. Kotlyar and D. Porath, Nano Lett. **7**, 981 (2007).

[11] C. Y. Ng, T. P. Chen, M. S. Tse, V. S. W. Lim, S. H. Y. Fung and A. A. Tseng, Appl. Phys. Lett. **86**, 152110 (2005).

[12] R. Dianoux, H. J. H. Smilde, F. Marchi, N. Buffet, P. Mur, F. Comin and J. Chevrier, Phys. Rev. B **71**, 125303 (2005).

[13] U. Zaghloul, B. Bhushan, P. Pons, G. J. Papaioannou, F. Coccetti and R. Plana, Nanotechnology **22**, 035705 (2011).

[14] S. Gómez–Moñivas, L. S. Froufe, R. Carminati, J. J. Greffet and J. J. Sáenz, Nanothecnology **12**, 496 (2001).

[15] F. Marchi, R, Dianoux, H. J. H. Smilde, P. Mur, F. Comin and J. Chevrier, J. Electrostat. **66**, 538 (2008).

[16] R. Dianoux, F. Martins, F. Marchi, C. Alandi, F. Comin and J. Chevrier, Phys. Rev. B **68**, 045403 (2003).

[17] S. Belaidi, P. Girard, and G. Leveque, J. Appl. Phys. **81**, 1023 (1997).

[18] T. D. Krauss and L. E. Brus, Phys. Rev. Lett. **83**, 4840 (1999).

[19]J. Jiang, T. D. Krauss and L. E. Brus, J. Phys. Chem. B **104**,11936 (2000).

[20]M. J. Gordon andT. Baron, Phys. Rev. B **72**, 165420 (2005).

[21]S. Belaidi, P. Girard and G. Leveque, J. Appl. Phys. **81**, 1023 (1997).

[22]L. Fumagalli, G. Ferrari, M. Sampietro, I. Casuso, E. Martıınez, J. Samitier and G. Gomila, Nanotechnology **17**, 4581 (2006).

[23]D. T. Lee, J. P. Pelz and B. Bhushan, Nanotechnology**17**, 1484 (2006).

[24]G. C. Qi, H. Yan, L. Guan, Y. L. Yang, X. H. Qiu, C. Wang, Y. B. Li and Y. P. Jiang, J. Appl. Phys. **103**, 114311 (2008).

[25]R. I. Revilla, X. J. Li, Y. L. Yang and C. Wang, J. Phys. Chem. C **118**, 5556 (2014).

[26]R. I. Revilla, Y. L. Yang and C. Wang, Surf. Interface Anal. **47**, 657 (2015).